\documentclass[trackchanges, twocolumn]{aastex701}

\usepackage{amsmath}            
\usepackage{amssymb}            %
\usepackage{amsfonts}           %
\usepackage{amsthm}             %

\usepackage{graphicx}           
\usepackage[utf8]{inputenc}     
\usepackage[nopatch=footnote]{microtype}
\usepackage{etoolbox}           
\usepackage[T1]{fontenc}        

\graphicspath{{}}       

\usepackage{anyfontsize} 
\usepackage{scalefnt}       
\newcommand\acro[1]{{\scalefont{.95}{#1}}} 

\renewcommand{\text}{\textnormal}	        
\renewcommand{\vec}[1]{\mathbf{#1}}       

\newcommand{\amber}{\textsc{amber}}
\newcommand{\pksz}{pk\acro{SZ}}
\newcommand{\hksz}{hk\acro{SZ}}
\newcommand{\popiiione}{Pop~\acro{III}.1}
\newcommand{\spt}{\acro{SPT-3G}}
\newcommand{\act}{\acro{ACT}}

\newcommand{\cmb}{\acro{CMB}}
\newcommand{\lcdm}{$\Lambda$\acro{CDM}}

\newcommand{\SO}[1]{\ifmmode
  \textnormal{\acro{SO(}}#1\textnormal{\acro{)}}
  \else \acro{SO($#1$)} \fi}

\newcommand{\SU}[1]{\ifmmode
  \textnormal{\acro{SU(}}#1\textnormal{\acro{)}}
  \else \acro{SU($#1$)} \fi}

\newcommand{\Sp}[1]{\ifmmode
  \textnormal{\acro{Sp(}}#1\textnormal{\acro{)}}
  \else \acro{Sp($#1$)} \fi}

\newcommand{\one}{\mathbbm{1}}

\newcommand{\D}[2][]{\ensuremath{\operatorname{d}\mkern-3mu^{#1}\mkern-1mu{#2}}\,}
 
\newcommand{\e}{\operatorname{e}}
\newcommand{\E}{\operatorname{e}}  

\newcommand{\I}{\operatorname{i}\mkern-2mu}  

\usepackage{scalerel} 

\usepackage{pifont}
\newcommand\Vtextvisiblespace[1][.3em]{%
  \mbox{\kern.06em\vrule height.3ex}%
  \vbox{\hrule width#1}%
  \hbox{\vrule height.3ex}}

\reportnum{\textcolor{gray!30!white}{\footnotesize\texttt{UCR-TR-2026-FLIP-03-K64}}}

\begin{document}

\title{Fireworks at Cosmic Dawn: relieving BAO--CMB tensions with the Pop III.1 Flash}

	\author[0000-0002-3862-0622]{Yash Aggarwal}
	\affiliation{Department of Physics and Astronomy, University of California, Riverside, CA 92521, USA}
	\email[show]{yagga003@ucr.edu}

	\author[0000-0001-9420-7384]{Christopher Cain}
	\affiliation{School of Earth and Space exploration, Arizona State University, Tempe, AZ 85281, USA}
	\email[show]{clcain3@asu.edu}

	\author[]{Garett Lopez}
	\affiliation{Department of Physics and Astronomy, University of California, Riverside, CA 92521, USA}
	\email{garettld@gmail.com}

	\author[0000-0001-6778-3861]{Hy Trac}
	\affiliation{McWilliams Center for Cosmology and Astrophysics, Department of Physics, Carnegie Mellon University, Pittsburgh, PA 15213, USA}
	\email{hytrac@andrew.cmu.edu }

	\author[]{Anson D'Aloisio}
	\affiliation{Department of Physics and Astronomy, University of California, Riverside, CA 92521, USA}
	\email{ansond@ucr.edu}

	\author[0000-0003-4642-2199]{Philip Tanedo}
	\affiliation{Department of Physics and Astronomy, University of California, Riverside, CA 92521, USA}
	\email{flip.tanedo@ucr.edu}

	\author[0000-0002-3389-9142]{Jonathan C. Tan}
	\affiliation{Department of Physics \& Astronomy, Chalmers University of Technology, Gothenburg, Sweden}
	\affiliation{Department of Astronomy \& Virginia Institute for Theoretical Astronomy, University of Virginia, Charlottesville, VA 22904, USA
	}
	\email{jctan.astro@gmail.com}

\begin{abstract}
A Cosmic Microwave Background (\acro{CMB}) optical depth of $\tau \sim 0.09$, several $\sigma$ in excess of the latest \textsl{Planck} low-$\ell$ \acro{EE} polarization measurement, has been proposed as a way to reconcile the preference for a sub-minimal neutrino mass sum in a combined analysis with \acro{CMB} and Dark Energy Spectroscopic Instrument (\acro{DESI}) three-year data. Reionization, however, is not just probed by $\tau$. It is also constrained by Ly$\alpha$ forest observations that indicate a late end of reionization, and the patchy kinetic Sunyaev-Zel'dovich (\pksz{}) effect which prefers a short duration. 
We explore whether an early phase of reionization can achieve a high $\tau$ while remaining consistent with both Ly$\alpha$ forest and \pksz{} constraints. As a concrete example, we consider supermassive Pop III.1 stars, dark-matter-powered metal-free stars proposed as progenitors of supermassive black holes. Within this framework, self-regulating ionizing feedback imposes a minimum source separation of $\sim 1 \, \text{cMpc}$, consequently limiting large-scale ionization fluctuations and reducing the \pksz{} power on observationally relevant scales. Our fiducial model realizes an optical depth of $\tau = 0.087$ with a Pop III.1-driven flash ionization phase centered at $z = 20$, while evading the most conservative $2\sigma$ upper limits on the \pksz{} signal from the most recent South Pole Telescope data release. More broadly, our findings motivate further exploration of early reionization models with weakly clustered sources as a possible resolution of tensions between \acro{BAO} and \acro{CMB} measurements.
\end{abstract}


\section{Introduction}

Recent measurements of baryon acoustic oscillations (\acro{BAO}) by the Dark Energy Spectroscopic Instrument (\acro{DESI}) have raised new possible tensions within the \lcdm{} framework~\citep{DESIDR22025}. In combination with cosmic microwave background (\cmb{}) data, these datasets favor sub-minimal neutrino masses, which formally extend to negative values when priors on neutrino mass ranges are removed~\citep[Fig.~15]{DESIDR22025}~\citep[see also][]{Craig2024,Loverde2024,Elbers2025a}.

Several solutions have been proposed, including dynamical dark energy~\citep{Elbers2025a, DESIDR22025}, new physics in the neutrino sector~\citep{Craig2024}, non-zero curvature~\citep{ChenZaldarriaga2025}, conversion of matter to dark energy~\citep{Ahlen2025}, and non-standard dark matter~\citep{KumarAjith2025}. We focus on a possibility that preserves \lcdm{} cosmology but increases the \cmb{} optical depth to reionization, $\tau \simeq 0.06 \rightarrow 0.09$~\citep{Sailer:2025lxj, Jhaveri:2025neg}, to a value several $\sigma$ larger than the latest \textsl{Planck} low-$\ell$ EE measurement~\citep{Planck2018,Tristram2024}.\footnote{This discrepancy may reflect instrumental systematics or model dependence in the inference of $\tau$.} A higher $\tau$ alleviates the neutrino mass tension by reducing $\Omega_\text{m}$ inferred from the \cmb{}, but requires reionization to be earlier and/or more extended than previously thought. The challenge, however, is that reionization is not just constrained by the integrated optical depth, but also by other observables that probe its timing, duration, and ionization morphology.

At face value, evidence of \emph{rapid} reionization emerges from the patchy kinetic Sunyaev-Zel'dovich (\pksz{}) effect: small scale Doppler shifts in the \cmb{} anisotropy spectrum, sourced by photons that Thomson scatter over free electrons with peculiar velocities~\citep{1970Ap&SS...7....3S,1991ApJ...372...21R,1995ApJ...455..419P} \citep[see also][]{Birkinshaw:1998qp}. Upper limits on the \pksz{} signal from the South Pole Telescope (\acro{SPT}) at $\ell=3000$ imply that the redshift interval between $25\%$ and $75\%$ ionization satisfies $\Delta z_{50}<4.1$~\citep{2021ApJ...908..199R}. Similar constraints are inferred from the four-point k\acro{SZ} statistic~\citep{Raghunathan2024}. Recent measurements from the Atacama Cosmology Telescope~\citep[\act{},][]{Beringue:2025bur} and \spt{}~\citep{SPT-3G:2026mcr} suggest even tighter limits.

At the same time, Ly$\alpha$ forest observations of high-redshift quasars indicate a \emph{late} end of reionization, $5.3 \lesssim z \lesssim 6$~\citep{Becker2015,Kulkarni2019,Keating2019,Nasir2020,Bosman2021,Becker2024,Zhu2024,Spina2024,Qin2024}. Measurements of the mean free path of ionizing photos further support this conclusion~\citep{Becker2021,Cain2021,Lewis2022,Zhu2023,Gaikwad2023}. Together, a late endpoint and short duration place strong limits on $\tau$. Indeed,~\citet{Cain:2025usc} show that ionization histories that end near $z=5.6$ and yield $\tau \simeq 0.09$ are in at least $\simeq 2\sigma$ tension with \pksz{} constraints. Subsequent analyses using Ly$\alpha$ forest data and Ly$\alpha$ transmission from high-redshift galaxies and quasars have reached similar conclusions~\citep{Elbers2025b,GarciaGallego2025,Kageura2026}. A key caveat is that these constraints usually assume smooth, monotonic reionization histories driven by galaxies.

Early Population III (Pop III) stars offer a qualitatively different possibility: a short-lived burst of ionizing photons at high redshift, followed by partial recombination and then a later, galaxy-driven reionization phase~\citep{Cen2003, Oh:2003pm, Furlanetto:2004nt, Ahn2012}. These ``double reionization'' scenarios were originally motivated by the high optical depths inferred from early WMAP measurements~\citep{2003ApJS..148..161K, 2003ApJS..148..175S, Komatsu2011}, but became less favored after \textsl{Planck} measured a lower value of $\tau$~\citep{Planck:2016mks,Wu2021}, though related models continue to be explored~\citep{Miranda2017,Ahn2021}.

An early Pop III-driven ionization phase may raise $\tau$ while evading \pksz{} limits for two reasons. First, the peculiar velocities that source the \pksz{} signal are smaller at higher redshift, which may be sufficiently small to offset enhancement from larger proper densities. Second, Pop III star formation is regulated by radiative and chemical feedback~\citep{Schauer2019,Visbal2020}. This means they form in minihalos that are less spatially biased than the galaxies that dominate later reionization. Less biased sources generate less clustered ionized structures which, in turn, suppresses the ionization fluctuations that contribute to \pksz{} power at $\ell\sim3000$~\citep{Park:2013mv,Lopez:2025bpg}.

One such realization is the \popiiione{} scenario---a model in which supermassive stars form in pristine, isolated minihalos~\citep[see review by][]{Tan:2024ysr} and are powered by annihilation of \acro{WIMP}-like dark matter particles~\citep{Spolyar:2007qv,2009ApJ...692..574N}. The \popiiione{} model is motivated to produce the cosmic population of supermassive black holes (\acro{SMBH}s), with most seeds forming at $z\sim20$--$30$~\citep{Banik:2016qww,Singh:2023beo,2025MNRAS.544.4317S,2026arXiv260528777P}. The supermassive star progenitors are also expected to produce a transient phase of ``flash ionization'' that can sufficiently raise $\tau$ during Cosmic Dawn~\citep{Tan:2025cua}, while remaining qualitatively consistent with \textsl{Planck} low-$\ell$ polarization measurements~\citep{Tan:2025obi}. The early phase of flash ionization alters the shape of the polarization spectra in a way that is distinct from ionization at lower redshifts~\citep[see, e.g.,][]{Zaldarriaga:1996ke}.

Dark-matter-powered protostars initially maintain large radii ($\sim 10^3 \, \mathrm{R}_\odot$) and cool photospheric temperatures ($\sim 10^4 \, \mathrm{K}$), weakening radiative feedback and sustaining high accretion rates. This allows the star to grow to masses $\gtrsim 10^4 \, \mathrm{M}_\odot$, far above the $10$--$100 \, \mathrm{M}_\odot$ range expected from conventional Pop \acro{III} formation channels~\citep[e.g.,][]{2008ApJ...681..771M} \citep[see also][]{Klessen:2023qmc}. 
After its dark matter fuel is exhausted, the star contracts toward the zero-age main sequence and emits an intense burst (flash) of ionizing radiation~\citep{Nandal:2025xvv,Topalakis:2025gux}. The H-ionizing photon luminosity may exceed $\sim 10^{53} \, \mathrm{s}^{-1}$ over a $\sim 10 \, \mathrm{Myr}$ lifetime, driving an R-type ionization front that generates ionized regions of order $\sim 1 \, \mathrm{cMpc}$~\citep{2025MNRAS.542.1532S,2026arXiv260528777P}. After the star dies, the ionized gas recombines on a timescale of $\sim 50 \, \mathrm{Myr}$, followed by the Pop \acro{II}-driven reionization phase.

The defining feature of the \popiiione{} flash-ionization framework is its self-regulation. Regions with enhanced ionization inhibit further \popiiione{} star formation by catalyzing $\mathrm{H}_2$ and $\mathrm{HD}$ formation, which promotes cooling and fragmentation in metal-free halos. These regions form significantly less massive and less luminous Pop \acro{III}.2 stars~\citep{2006MNRAS.373..128G} \citep[see also][Ch.~5.3]{2013fgu..book.....L}. Therefore, the feedback radius sets the abundance of ionizing sources and \acro{SMBH}s. The cosmological impact of \popiiione{} flash ionization can be effectively parameterized by the isolation distance $d_\mathrm{iso}$~\citep{Banik:2016qww,Singh:2023beo,2025MNRAS.536..851C}. More recently, \citet{2026arXiv260528777P} presented a version of the model in which self-regulated feedback from \popiiione{} stars yields a fiducial \acro{SMBH} abundance of $n_{\rm SMBH}\sim 0.1 \, \text{cMpc}^{-3}$.

By regulating the growth and merging of ionized regions, the \popiiione{} framework may suppress the fluctuations in the ionized momentum that source \pksz{} power on observationally relevant scales, $\ell \sim 3000$. During cosmic dawn these poles correspond to length scales of $\sim 20 \, \text{cMpc}$, much larger than the $d_\text{iso}$ scale. We investigate this possibility as follows. \S\ref{sec:flash:ionization:model} describes the evolution of the globally averaged ionization fraction in the flash ionization scenario. \S\ref{sec:methodology} describes simulation setup and source seeding prescriptions used to explore clustering and isolation effects. Results are discussed in \S\ref{sec:results}, 
and we conclude in \S\ref{sec:conclusion:discussion}.  We adopt a flat $\Lambda\text{CDM}$ cosmology with parameters as follows: $h=0.68$, $\Omega_m = 0.305$, $\Omega_\Lambda = 0.695$, $\Omega_b = 0.048$, $\sigma_8 = 0.82$, and $n_s = 0.9667$.

\section{Pop III.1 Flash Ionization Model}
\label{sec:flash:ionization:model}

We adopt a flash ionization history from~\citet{Tan:2025cua, Tan:2025obi} and combine it with a baseline Pop \acro{II} only reionization model. The ionization fraction during the \popiiione{} flash is modeled as a linear function of cosmic time $t$, while the subsequent recombination phase is modeled as an exponential decay
\begin{align}
    \bar{x}_\text{III.1} (z)
    & = 
    \begin{cases}
        \bar{x}_\text{max} \frac{t - t_\text{on}}{t_\text{flash} - t_\text{on}} 
        & z_\text{on} \geq z > z_\text{flash}  \\ 
        \bar{x}_\text{max} \exp\left({-\frac{t - t_\text{flash}}{t_\text{rec}}}\right)
        & z_\text{flash} \geq z > z_\text{off} 
    \end{cases}
    \ .
    \label{eq:iii.1:ion:history}
\end{align}
where $\bar{x}_\text{max}$ is the maximum ionization fraction at $z_\text{flash}$, $z_\text{on}$ and $z_\text{off}$ describe the duration of \popiiione{} phase (see \S\ref{sec:simulation:setup}), and the modified recombination time scale $t_\text{rec}$ is
\begin{align}
    t_\text{rec} 
    & \approx
    55 \, \text{Myr}
    \left( \frac{1+z_\text{flash}}{21} \right)^{-3} \ ,
\end{align}
assuming an IGM clumping factor of $3$. This factor approximates the effect of gas cooling via Thomson scattering over \cmb{} photons and Hubble expansion in the post-flash period. At lower redshifts, reionization is driven by Pop \acro{II} stars residing in galaxies. We adopt a baseline history $\bar{x}_\text{gal}(z)$ parameterized by midpoint $z_\text{mid} = 7.17$, duration $\Delta z = 5.8$, and asymmetry $A_z = 3.0$, implemented using the Weibull form of~\citet{Trac:2021qbn}.  This baseline is consistent with both Ly$\alpha$ forest and \pksz{} constraints~\citep{Cain:2025usc}.

\begin{figure}[t]
    \centering
    \includegraphics[width=\linewidth]{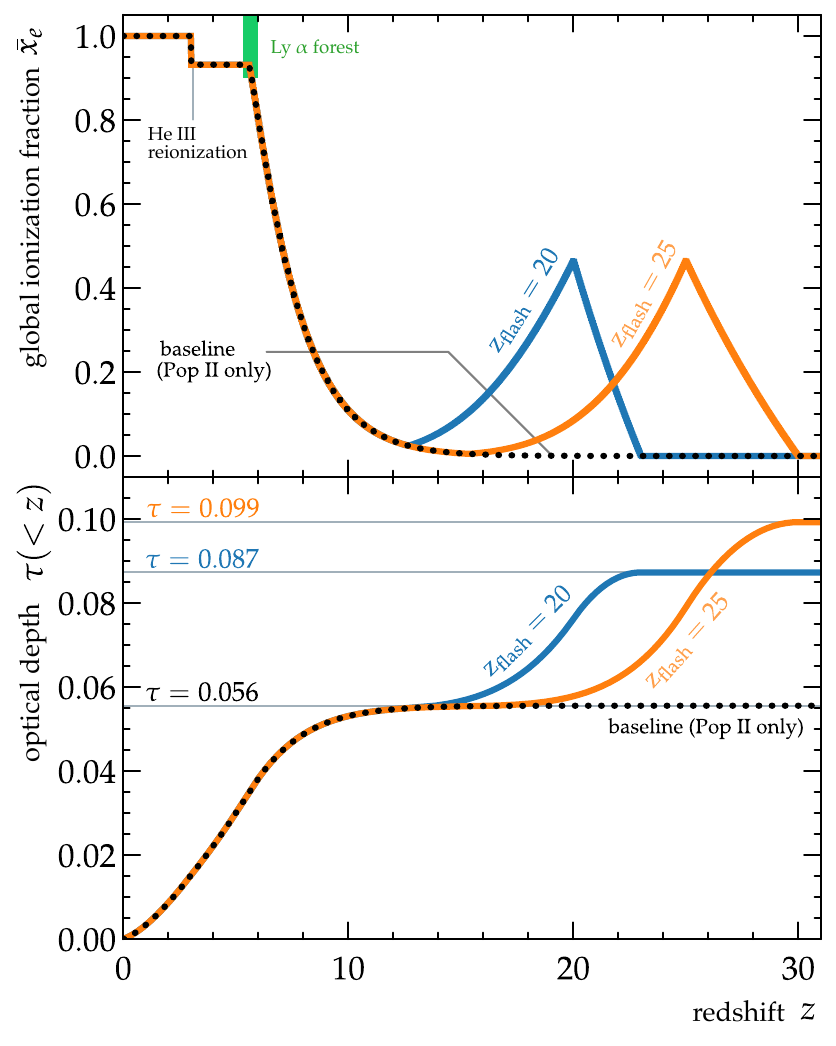}
    \caption{Model ionization histories used in our \amber{} runs. \emph{Top:} globally averaged, mass-weighted free electron fraction, defined to be unity when Hydrogen and Helium are fully ionized. We assume that Helium is singly ionized for $z > 3$ and doubly ionized for $z\leq 3$. The green region denotes required end of reionization from the Ly$\alpha$ forest.
    \emph{Bottom:} Corresponding cumulative CMB optical depths.}
    \label{fig:reionization:history}
\end{figure}

Because the two ionization episodes are separated by significant cosmic time, we may model the full reionization history as 
\begin{align}
    \bar{x}_e (z)
    & = 
    \text{max}
    \left[ \bar{x}_\text{gal} (z), \bar{x}_\text{III.1} (z) \right] \ . 
    \label{eq:reionization:history}
\end{align}
The gas nearly recombines by the end of the \popiiione{} phase ($z \gtrsim z_\text{off}$) so that the subsequent galaxy-driven phase is mostly independent of the flash. Figure~\ref{fig:reionization:history} shows the reionization histories used in this work and their corresponding optical depths.

\section{Methodology}
\label{sec:methodology}

\subsection{Simulation Setup}
\label{sec:simulation:setup}

We modify the semi-numerical code Abundance Matching for the Epoch of Reionization (\textsc{amber})~\citep{Trac:2021qbn,Chen:2022lhr} to simulate \popiiione{} flash ionization and model its \pksz{} signal. Our simulation uses box sizes of $ L_\text{box} = 250 \, \text{cMpc}/h$ with $1024$ cells per side, yielding a spatial resolution of $L_\text{cell} = 0.244 \, \text{cMpc}/h$. For each parameter configuration in our study, we generate an ensemble of 27 independent \amber{} realizations and report the mean \pksz{} power as our primary prediction.

During the \popiiione{} phase, \textsc{amber} initializes all sources at $z = z_\text{on}$ according to one of the stochastic seeding prescriptions described below in \S\ref{sec:seeding:models}. Our models are characterized by a global number density and, where applicable, an isolation distance. We assume fiducial values for these parameters,
\begin{align}
    \bar{n}_\text{III.1} &= 0.1 \, \text{cMpc}^{-3}
    & 
    d_\text{iso} &= 1.55 \, \text{cMpc} \ .
    \label{eq:fiducial:iii1:parameters}
\end{align}

\amber{} employs $d_\text{iso}$ in two distinct ways: to set the minimum separation between any two sources, and to set the mean free path in \amber{}'s radiation kernel, $\lambda_\text{mfp} = d_\text{iso}$ (see~\citet[][]{Trac:2021qbn} for details on the radiation kernel). Together, these modifications enforce the isolation scale and ensure that the sizes of ionized bubbles, which are controlled by $\lambda_{\rm mfp}$, are consistent with $d_{\rm iso}$.

The \popiiione{} phase consists of a flash ionization stage followed by gas recombination. During the flash stage, $z_\text{on} \geq z \geq z_\text{flash}$, ionized regions expand and merge until the global ionization fraction reaches $\bar{x}_\text{max}$ at $z_\text{flash}$, with each cell treated as either ionized or neutral. For $z_\text{flash} > z > z_\text{off}$, the ionization morphology is held fixed at its $z_\text{flash}$ configuration, while the ionized fraction is uniformly reduced to model recombination. This approximation encodes recombination through the prescribed global ionization history, equation~\eqref{eq:reionization:history}. Appendix~\ref{sec:amber:changes} provides further technical details about how the Pop III.1 reionization history is implemented in \textsc{amber}. We set $z_\text{off} = 12.25$ for all flash ionization runs, corresponding to redshift at which the global ionization fraction reaches its minimum value in the $z_\text{flash}=20$ model. The resulting \pksz{} power generated during the \popiiione{} phase is not sensitive to this choice, since recombination is largely complete by this redshift. Fixing $z_\text{off}$ in this way also sets the onset of the subsequent galaxy-driven phase, whose contribution is consistent with the baseline galaxy-driven reionization history. Below $z \leq z_\text{off}$, reionization is driven by galaxies residing in $M \geq 10^8 h^{-1} \, \text{M}_\odot$ halos with $\lambda_\text{mfp} = 3.0 h^{-1} \, \text{cMpc}$, following the standard procedure in \citet{Trac:2021qbn, Chen:2022lhr}.

\begin{figure*}[!t]
    \centering
    \includegraphics[width=\textwidth]{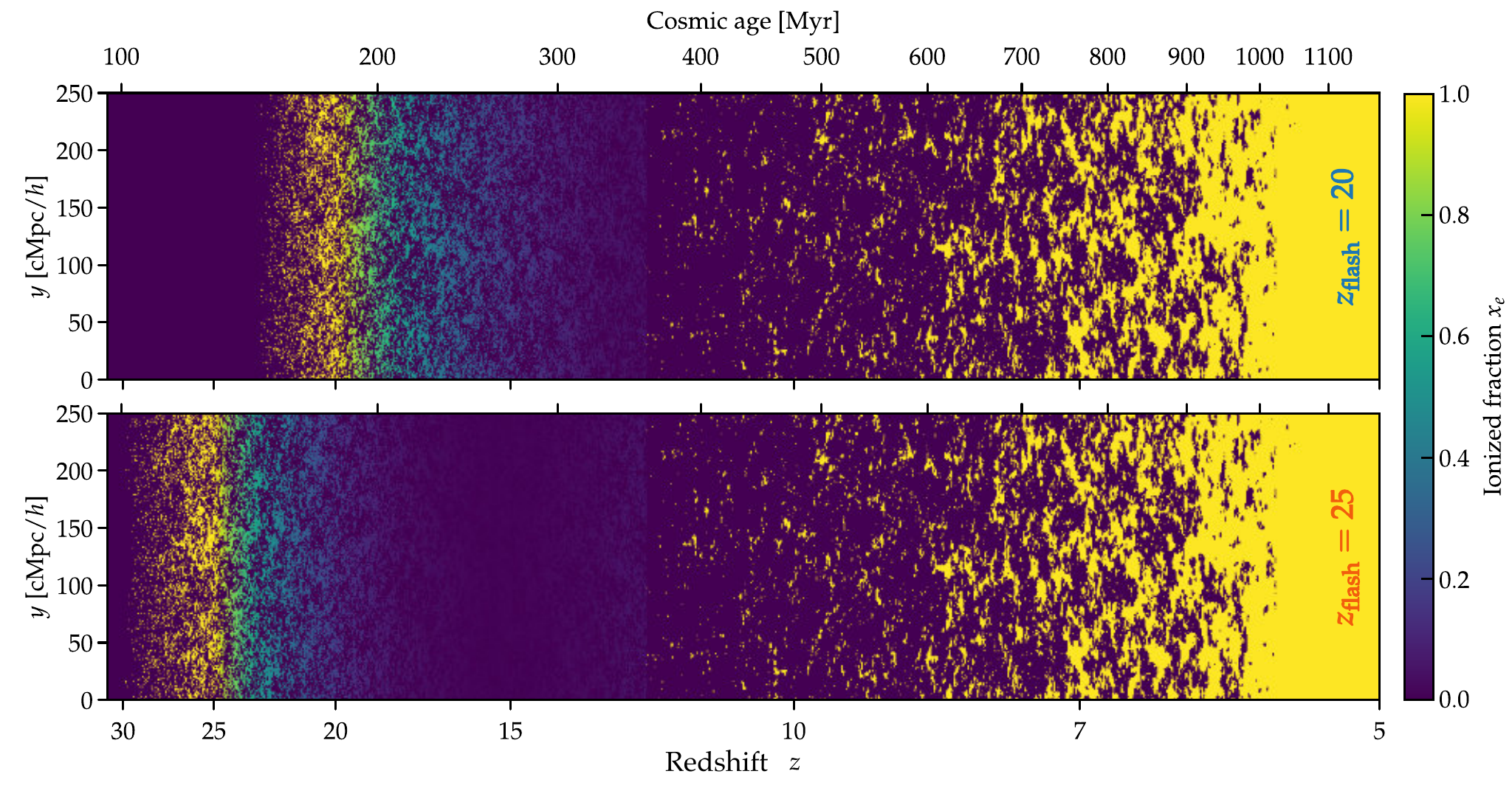}
    \caption{Time evolution of the ionization field for the $z_{\rm flash}=20$ (top) and $z_{\rm flash}=25$ (bottom) models using our fiducial \popiiione{} parameters and the Biased Exclusion seeding prescription (see \S \ref{sec:seeding:models}). The figure shows the characteristic behavior and scale of the ionization field during the three stages of evolution in our simulation: initial \popiiione{} flash phase, followed by uniform recombination, and a later reionization driven by galaxies.}
    \label{fig:time:slices}
\end{figure*}
Figure~\ref{fig:time:slices} visualizes the ionization field along a lightcone for the model flash ionization runs using the Biased Exclusion seeding prescription (see \S\ref{sec:biased:exclusion}) and our fiducial \popiiione{} parameters in equation~\eqref{eq:fiducial:iii1:parameters}. This figure highlights the characteristic behavior of the ionization field for each epoch: during the flash stage ionized regions grow and merge until $z_{\rm flash}$.  After this, the ionization morphology is ``frozen'', while gas recombines uniformly.  At lower redshifts, Pop II galaxies take over and drive reionization.

We compute the \pksz{} power spectrum using the Limber approximation following~\citet[Eq.~6]{Park:2013mv}, and report it as $D_\ell \equiv \ell(\ell+1)C_\ell/(2\pi)$. Since our relatively small simulation volumes miss part of the large-scale velocity contribution to the ionized momentum field, corresponding to modes with $k < k_\text{box}$, we correct for these missing modes using the prescription of~\citet[Appendix B]{Park:2013mv}; see also~\citet{Lopez:2025bpg}. Furthermore, the box size is limited by the need to resolve the Pop~III.1 isolation scale $d_\text{iso}$, which governs the morphology of the ionization field. The finite simulation volume introduces realization-to-realization scatter, so for each parameter configuration we run $27$ independent \amber{} realizations. The mean \pksz{} power and corresponding $1\sigma$ scatter for our benchmark models are presented in Appendix~\ref{sec:ensemble:convergence}.

An important caveat of our work is that we do not model the \popiiione{} framework fully self-consistently. In a complete treatment, the isolation scale $d_\text{iso}$ would regulate the formation and spatial distribution of \popiiione{} sources. Instead our \popiiione{} parameter space is motivated by models of supermassive black hole formation~\citep{Banik:2016qww,Singh:2023beo, Tan:2024ysr}. Likewise, the flash ionization histories used in our analysis are adopted from~\citet{Tan:2025cua,Tan:2025obi}, rather than computed self-consistently from the evolving source population. Thus, our analysis should be interpreted as a controlled phenomenological study of how a minimum isolation constraint, source abundance, and flash ionization history affect the resulting \pksz{} signal, rather than as a fully self-consistent model of \popiiione{} star formation.

\subsection{Pop-III.1 seeding models}
\label{sec:seeding:models}

The \pksz{} signal depends on the morphology of ionized regions, which is set by the spatial distribution of ionizing sources. We model the source field as the number of \popiiione{} stars, $ N_i \equiv N_\text{III.1} (\vec{r}_i)$, in cell at location $\vec{r}_i$.\footnote{The overall normalization is unimportant in \amber{} since ionized regions are determined by relative ranks of sources and the prescribed global ionization history~\citep[see][Sec.~3.4.1]{Trac:2021qbn}.} The sources are populated using three stochastic seeding prescriptions: \emph{Biased Poisson}, \emph{Biased Exclusion}, and \emph{Random Exclusion}, described below. In all cases, sources are seeded instantaneously at $z = z_\text{on}$, after which the global ionization field evolves according to \S\ref{sec:flash:ionization:model}.  

\begin{figure*}[!t]
    \centering
    \includegraphics[width=\textwidth]{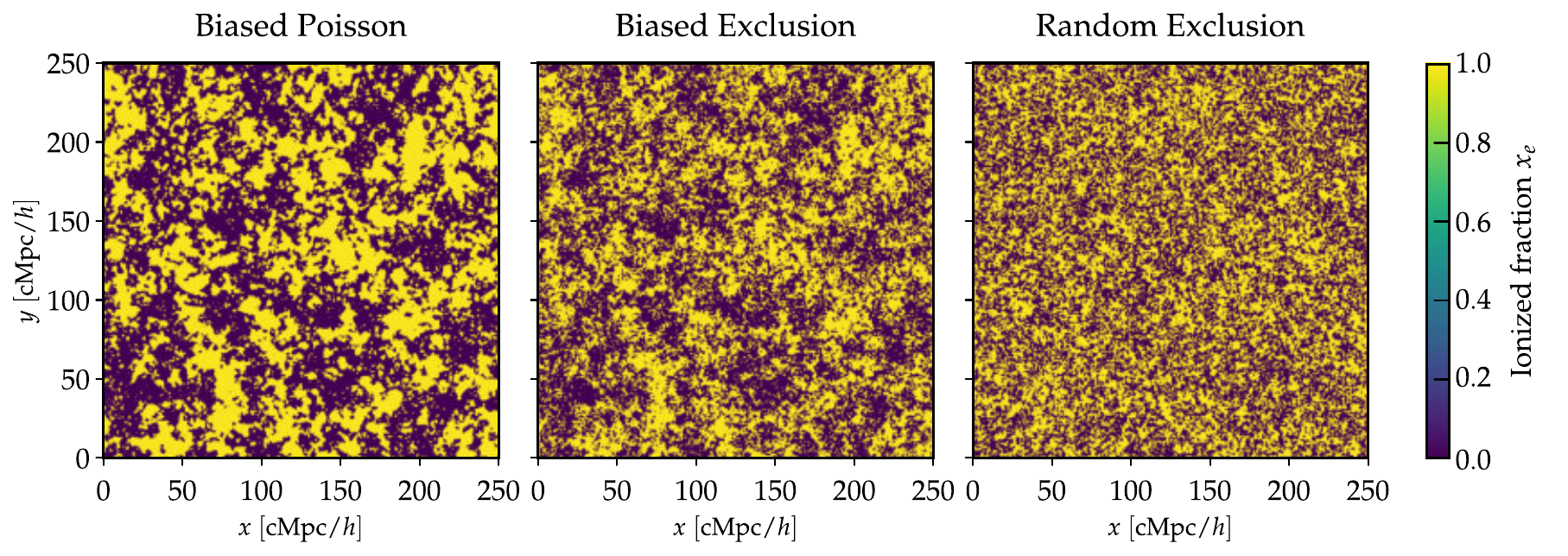}
    \caption{Ionization field slice at peak ionization fraction for the $z_\text{flash}=20$ run, shown for three seeding prescriptions: Biased Poisson (\emph{left}), Biased Exclusion (\emph{middle}), and Random Exclusion (\emph{right}), using fiducial \popiiione{} parameters in equation~\eqref{eq:reionization:history}. See \S\ref{sec:effect:seed:model} for further details.}
    \label{fig:ion:morphology:flash20}
\end{figure*}

\subsubsection{Biased Poisson}
\label{sec:biased:Poisson}

In this model, \popiiione{} stars are drawn from an inhomogeneous Poisson distribution~\citep[Ch.~5]{ross2014introduction}, with local mean
\begin{align}
    \mu (\vec{r}_i) 
    & =
    \bar{n}_\text{III.1} L_\text{cell}^3
    \frac{\Delta f (\vec{r}_i)}{\langle \Delta f \rangle}
    \ ,
\end{align}
where
\begin{align}
    \Delta f (\vec{r}_i)
    & = 
    f (\geq M_\text{low} | \vec{r}_i, z_\text{on}) - 
    f (\geq M_\text{high} | \vec{r}_i, z_\text{on}) \ . 
\end{align}
$f(\geq M | \vec{r}_i, z)$ is the collapsed mass fraction in halos with mass above $M$ at cell $\vec{r}_i$ and $z$, and $\langle \cdot \rangle$ averages over the simulation volume. We adopt $M_\text{low} = 5\times 10^5 \, \text{M}_\odot$ and $M_\text{high} = 5 \times 10^6 \, \text{M}_\odot$ for the candidate \popiiione{} host-halo range. The total source count has expectation value $\sum_i \mu(\vec{r}_i) = \bar{n}_\mathrm{III.1} L_\mathrm{box}^3$ and Poisson standard deviation $\sqrt{\bar{n}_\mathrm{III.1} L_\mathrm{box}^3}$.

This model represents a purely Biased Poisson sampling of the halo field, with clustering inherited from $\Delta f$ and no regional suppression due to radiative feedback.

\subsubsection{Biased Exclusion}
\label{sec:biased:exclusion}
This model extends the \emph{Biased Poisson} prescription by imposing spatial exclusion. The population is biased by the collapsed fraction field while being spatially regulated by the exclusion scale $d_\mathrm{iso}$. Cells are ranked by formation time, $t_i$. A \popiiione{} source is accepted ($N_i = 1$) only if it lies farther than $d_\mathrm{iso}$ from all previously accepted sources; otherwise, it is discarded. The process continues until $\bar{n}_\mathrm{III.1}L_\mathrm{box}^3$ sources are placed.

Each cell at $\vec{r}_i$ draws a stochastic formation time $t_i$ from an exponential distribution~\citep[Ch.~5]{ross2014introduction}, with local formation rate proportional to $f(\geq 10^6 \, \mathrm{M}_\odot | \vec{r}_i,z_\mathrm{on})$.\footnote{Mathematically, this procedure corresponds to first-arrival problem for an inhomogeneous Poisson process.} The normalization of the rate is irrelevant since only the rank ordering of $t_i$ is used to populate sources. This prescription ranks \emph{which cells form stars first}, rather than how many stars form per cell. We use the cumulative collapsed fraction $f(\geq 10^6 \, \mathrm{M}_\odot)$, rather than the binned fraction used in the Biased Poisson model, to identify regions associated with the earliest formation events.

\subsubsection{Random Exclusion}
\label{subsubsec:random}
\popiiione{} sources are placed randomly throughout the simulation volume, independent of the underlying density field. A hard exclusion region is imposed such that no two sources lie within a comoving distance $d_\text{iso}$.
This model therefore represents a spatially-regulated distribution of sources with no correlation to the collapsed fraction field.  

\subsubsection{Impact of seeding prescriptions}
\label{subsubsec:impact}

Figure~\ref{fig:ion:morphology:flash20} shows the morphology of the ionization field at peak flash ionization for the $z_\text{flash} = 20$ run with fiducial \popiiione{} parameters. The Biased Poisson model shows the strongest ionization clustering, and is this expected to have the highest pkSZ power at $\ell = 3000$. The Biased Exclusion model mitigates the effect of source clustering by imposing a minimum separation $d_\text{iso}$ and suppressing radiation beyond this exclusion region. The Random Exclusion model is uncorrelated with the underlying density field and should contribute least to the \pksz{} power.

\section{Results}
\label{sec:results}
\begin{figure*}[t!]
    \centering
    \includegraphics[width=\textwidth]{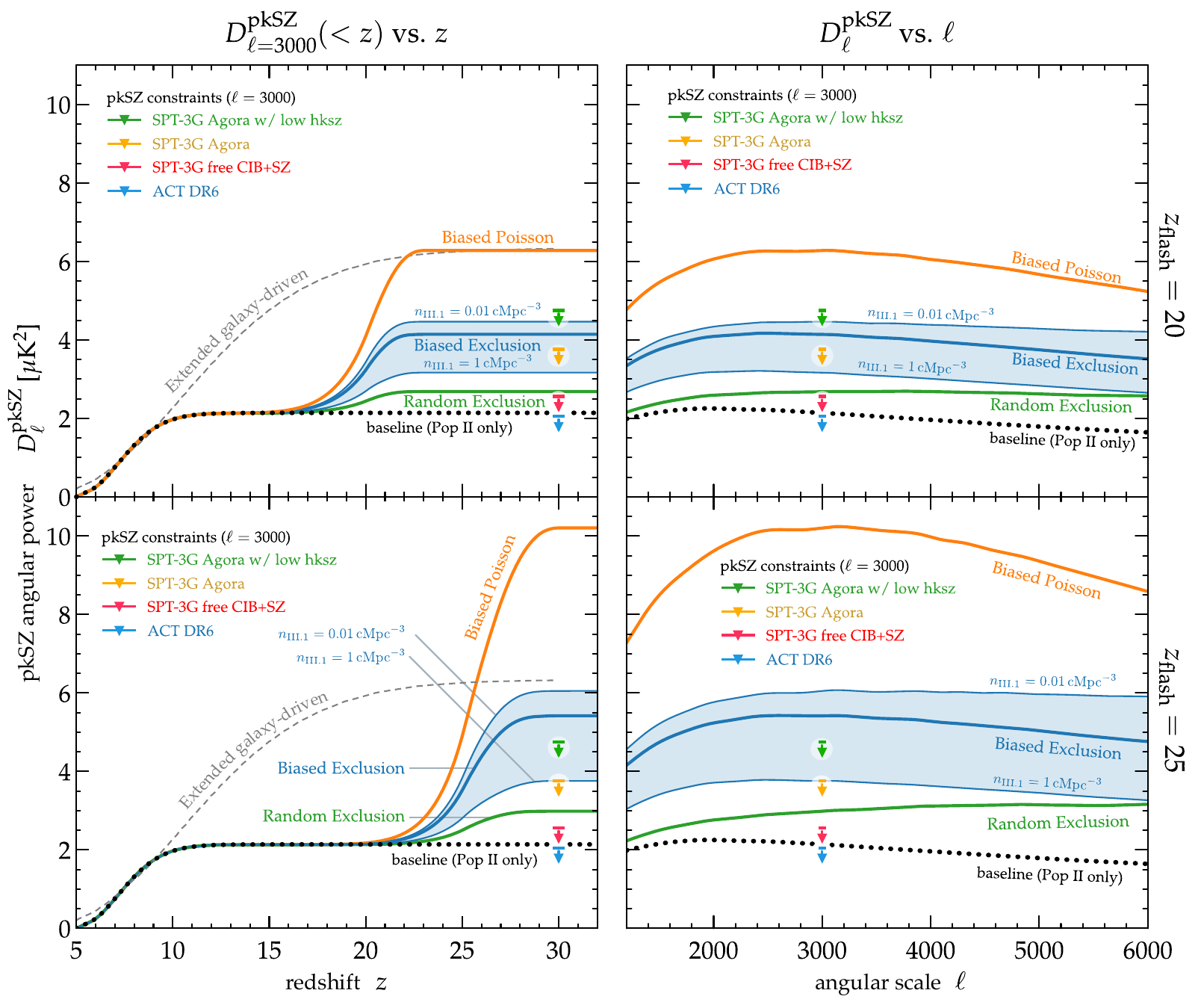}
    \caption{
    \pksz{} power for \popiiione{} flash ionization models with fiducial parameters in equation~\eqref{eq:fiducial:iii1:parameters},
    including the contribution from baseline galaxy-driven model. The left panels show the cumulative contribution $D_{\ell=3000}^\text{pkSZ}(<z)$, while the right panels show the full angular power spectrum $D_\ell^\text{pkSZ}$. The top and bottom rows correspond to $z_\text{flash}=20$ and $25$ ionization histories, respectively (Fig.~\ref{fig:reionization:history}).
    Colored curves denote results from different seeding prescriptions described in \S\ref{sec:seeding:models}, while the dotted black curves show the results of the baseline galaxy-driven reionization model. The shaded region brackets a one dex variation in \popiiione{} number density for the Biased Exclusion model 
    with $n_\text{III.1}$ given on the plot and $d_\text{iso}$ related by equation~\eqref{eq:niii:diso}.
    The arrows show the upper $2 \sigma$ limits on \pksz{} power from \act{}~\citep{AtacamaCosmologyTelescope:2025blo,Beringue:2025bur} and \spt{}~\citep{SPT-3G:2026mcr}. 
    The dashed grey curves in the left panels show the \pksz{} power from an extended galaxy-driven reionization model from~\citet[dashed-red Fig.~3]{Cain:2025usc} which satisfies the citeria of a late end to reionization ($ z \lesssim 6$) from Ly$\alpha$ forest measurements and a high $\tau \sim 0.09$, but exceeds limits on the \pksz{}. 
    }
    \label{fig:results}
\end{figure*}

\popiiione{} flash ionization introduces an early-time contribution to the \pksz{} that is sensitive to the assumed source clustering. Figure~\ref{fig:results} shows the \pksz{} angular power spectra for our fiducial \popiiione{} parameters shown in equation~\eqref{eq:fiducial:iii1:parameters}. The left column displays the cumulative contribution $D_{\ell=3000}^\text{pkSZ}(<z)$, while the right column shows the full angular power spectrum $D_\ell^\text{pkSZ}$, for each seeding model (see legend) and flash ionization histories (different rows). The shaded region brackets the signal for the Biased Exclusion model with one dex variation in the source number density, with isolation distance scaled according to~\citep[Eq.~9]{Tan:2024ysr} 
\begin{align}
    \bar{n}_\text{III.1} d_\text{iso}^3 &= \text{const}
    \label{eq:niii:diso}
\end{align}
Higher (lower) number densities give lower (higher) \pksz{} amplitudes.

\subsection{\texorpdfstring{Limits on the \pksz{} signal}{Limits on the pkSZ signal}}
Data points indicate the $2\sigma$ upper limits on the allowed $D_{\ell=3000}^\text{pkSZ}$ signal from \cmb{} observations by the \act{}~\citep{AtacamaCosmologyTelescope:2025blo,Beringue:2025bur} and \spt{}~\citep{SPT-3G:2026mcr} collaborations. The recent \spt{} data release demonstrates that the inferred total kSZ signal depends sensitively on foreground modeling, with the thermal SZ $\times$ \acro{CIB} contribution being the most significant source of uncertainty. We show the results for the \acro{AGORA} and free \acro{CIB}$\times$\acro{SZ} fitting templates~\citep[Table~5]{SPT-3G:2026mcr}.

Since observations measure the total \acro{SZ} signal, homogeneous k\acro{SZ} (\hksz{}) + \pksz{}, we must subtract an estimate for the hk\acro{SZ} contribution to recover a \pksz{} constraint. The template cases for \spt{} use a relatively high estimate for the \hksz{} contribution, $D_{\ell=3000}^\text{hkSZ} = 1.84 \, \mu\text{K}^2$, obtained from \citet[Eq.~8.1]{SPT-3G:2026mcr}.  The ``\act{}'' and ``\spt{} \acro{AGORA} w/ low \hksz{}'' constraints use a low, $D_{\ell=3000}^\text{hkSZ} = 0.85 \, \mu\text{K}^2$, \hksz{} contribution estimate from \acro{AGORA} simulations~\citep{Omori:2022uox}. The ``\spt{} \acro{AGORA} w/ low \hksz{}'' data point represents the most conservative upper limit, which uses both the highest limit on the total k\acro{SZ} (from \spt{}) and the lowest estimate for the \hksz{}.

\subsection{Dependence on the seeding prescription}
\label{sec:effect:seed:model}

The imprint of flash ionization on the \pksz{} signal in our Biased Exclusion model can be understood from the characteristic scales involved. The comoving distance between ionized regions at $z \sim 20$--$30$ and the present universe is of order $s \sim 11 \, \text{cGpc}$ so that $\ell \sim 3000$ probes fluctuations on scales of order $\lambda_\text{char} \sim \left(\frac{2 \pi}{\ell}\right)s \sim 23 \, \text{cMpc}$. However, in the \popiiione{} framework ionization fluctuations are expected to occur on scales of order $d_\text{iso}$, as this determines source clustering and typical size of ionized regions. Since $d_\text{iso} \ll \lambda_\text{char}$, the resulting ionization structure is shifted to smaller scales than those probed at $\ell \sim 3000$, suppressing the \pksz{} signal on observationally relevant scales.  
In the Random Exclusion model, the absence of correlations between the source and density fields further suppresses clustering, resulting in almost no contribution to the \pksz{} signal from the Pop III.1 phase. In the Biased Poisson model, the opposite is true---the lack of exclusion increases ionization fluctuations at $\sim \lambda_{\rm char}$, resulting in a \pksz{} signal comparable to Pop II-only models.  The relative contribution of each seeding model to the \pksz{} can be understood from Figure~\ref{fig:ion:morphology:flash20}---models with more biased ionization fluctuations produce larger \pksz{} signals.

The right panels of Figure~\ref{fig:results} show that the angular dependence of the \pksz{} power is also sensitive to the seeding model. In addition to variations in total power, there are subtle shape differences between the different source models. Scenarios with less biased source clustering have smaller ionization fluctuation at large scales and more fluctuations at small scales, which pushes power from low to high $\ell$.  We thus speculate that measurements of the \pksz{} at much higher $\ell$ than $3000$ could help distinguish these different source clustering scenarios, echoing the findings of~\citet{Lopez:2025bpg}.

The shaded region brackets the \pksz{} power for a one dex variation in the source number density in the Biased Exclusion model. The power decreases with increasing source number density and/or isolation distance. Increasing the isolation distance increases the spatial separation among sources, whereas increasing the number density, while keeping the reionization model fixed, decreases their ionizing luminosity. Both effects localize ionized regions around individual sources, suppressing large scale clustering and merging.

\subsection{\texorpdfstring{Consistency with \pksz{} limits}{Consistency with pkSZ limits}}
\label{subsec:consistency}
We see that the $z_\text{flash}=20$ model predicts a lower \pksz{} than the $z_\text{flash}=25$ case, and is thus in better accord with upper limits.  This is consistent with the findings of~\cite{Tan:2025obi} as it pertains to the low-$\ell$ EE power spectrum.  As such, we will focus on the Flash 20 case in this section.

The Biased Poisson model predicts a \pksz{} signal similar to the reference high-$\tau$ model from~\cite{Cain:2025usc}, and is thus inconsistent with all the $2\sigma$ upper limits we show.  In this model, the clustering of Pop III.1 sources is similar to that of Pop II galaxies at lower redshifts, resulting in a large contribution to the signal. The fiducial Biased Exclusion model, however, predicts a \pksz{} power of $\approx 2$ $\mu$K$^{2}$, bringing it into agreement with the ``\spt{} \acro{AGORA} w/ low \hksz{}'' limit. Assuming a Pop III.1 number density an order of magnitude than our fiducial value 
\eqref{eq:fiducial:iii1:parameters} also agrees with the ``\spt{} \acro{AGORA}'' limit, which does not require the hkSZ component to be on the low end of theory predictions.

The Random Exclusion model, which can be understood as as limiting case of feedback-driven source de-biasing, evades both of these limits and almost reaches the ``\spt{} \acro{AGORA} free CIB $\times$ tSZ'' limit. The upper limit from ACT is low enough that even the signal from the baseline Pop II-driven reionization model is in mild tension with it.  

We see that Pop III.1 reionization models can, under certain conditions, produce \pksz{} signals in agreement with recent upper limits from SPT.  This is in contrast with standard Pop II-driven reionization models and Pop III models in which sources tightly trace the density field without any feedback effects.  More generally, our findings suggest that any Pop III reionization model that shares the Pop III.1 feature of large-scale ($\sim {\rm cMpc}$) feedback-driven exclusion could be a candidate to achieve a similar result.

\section{Conclusion and Discussion}
\label{sec:conclusion:discussion}

We have modeled the imprint of \popiiione{} flash ionization on the \pksz{} signal in the \cmb{}, generated during Cosmic Dawn. 
Our fiducial model, in which feedback effects limit the clustering of \popiiione{} sources, is able to evade current upper limits on the \pksz{} signal whilst concluding reionization at $z < 6$, in accord with the Ly$\alpha$ forest.  Further, our fiducial model predicts $\tau = 0.087$, in agreement with the values required to resolve the \acro{BAO}-\acro{CMB} neutrino mass tension \citep{Sailer:2025lxj, Jhaveri:2025neg}.  
Our main findings can be summarized as follows:
\begin{itemize}

    \item An early ``flash'' phase of reionization driven by Pop III.1 stars can realize a large value of $\tau \sim 0.09$ without requiring an early end to reionization. Unlike scenarios where Pop II galaxies drive reionization, Pop III.1 stars may ease tensions between the CMB and recent DESI data in a way that can be reconciled with Ly$\alpha$ forest and \pksz{} constraints.  Our preferred model  has $\tau = 0.087$ and stays under the most conservative \pksz{} upper limits from SPT-3G.  As such, Pop III.1 stars may present an observationally viable pathway to resolving the neutrino mass tension with high $\tau$.  
    
    \item The feedback-driven clustering of Pop III.1 sources drives our main result.  Scenarios in which feedback imposes a minimum separation between adjacent Pop III.1 sources have substantially less \pksz{} power than those in which Pop III.1 clustering is assumed to trace the density field.  Our findings establish that \emph{any} Pop III reionization model that predicts minimal large-scale ionization fluctuations at high redshift may be able to resolve this tension in the same manner.  

    \item Our fiducial model satisfies the $2\sigma$ limits from SPT-3G that (1) use simulation-based templates for the CIB $\times$ tSZ cross-correlation  {\it and} the (2) assume that the low-redshift contribution to the total kSZ is on the low end of theoretical expectations. Relaxing the second assumption while maintaining $2\sigma$ consistency requires a higher Pop III.1 number density than our fiducial value \eqref{eq:fiducial:iii1:parameters}. Upper limits from SPT-3G that assume a free-form template for the CIB $\times$ tSZ and recent upper limits from ACT are in mild tension with our baseline Pop-II-driven reionization model, without any Pop III.1 contribution.  Establishing consistency with these limits would likely require a later, more rapid Pop II-driven phase and a much higher number density of Pop III.1 sources.
\end{itemize}

Our findings motivate further investigation of models in which Pop III stars contribute significantly to the early phases of reionization. \acro{JWST} observations may have already begun to reveal candidate Pop III sources at lower redshift~\citep{Fujimoto2025,Venditti2025}.  Observations of Ly$\alpha$ emitters and galaxy damping wings have pushed constraints on the early phases of reionization above $z = 10$~\citep{Umeda2023,Witstok2024,Qin2024,Cohon2026,2026ApJ...997...86U}, edging closer to a possible Pop III-driven phase.  Forthcoming observations have the potential to support or disfavor this possible resolution to the \acro{BAO}-\cmb{} tension.


\bigskip
\section*{Acknowledgment}

The authors thank Kevin Croker for helpful comments on the draft version of this manuscript.  The \amber{} modifications and individual runs in this work were performed using the standard \textsc{compute} nodes on the Expanse system at San Diego Supercomputer Center~\citep{10.1145/3437359.3465588}. The ensemble runs were performed using the \textsc{ICX} and \textsc{PVC} nodes on Stampede3 system at the Texas Advanced Computing Center (\acro{TACC}), URL: http:\url{www.tacc.utexas.edu}. AD acknowledges support from NSF grant AST-2045600.  CC acknowledges support from the Beus Center for Cosmic Foundations at Arizona State University.

\appendix

\section{Modeling non-monotonic reionization histories in AMBER}
\label{sec:amber:changes}

In this section, we describe our modifications to the publicly available version of \amber{}.  In its original form, the code constructs a continuous reionization-redshift field $z_\text{re}(\vec{r})$ by abundance matching at a single reference redshift. Cells are ranked according to the intensity of the local radiation field, and the cumulative mass fraction of the most highly illuminated regions is matched to the globally averaged ionization fraction. This approach implicitly assumes a monotonic ionization history, in which once a region is ionized it remains ionized thereafter.

To include flash ionization models, we split \amber{} into three phases: \popiiione{} flash, recombination, and galaxy-driven. The ionization history rises monotonically with cosmic time during the \popiiione{} ($z_\text{flash} \leq z \leq z_\text{on}$) and galaxy-driven ($z \leq z_\text{off}$) stages. Consequently, we separate the ionization histories in these stages and use \amber{}'s native abundance matching prescription to create two independent sets of reionization redshift fields---$z_\text{re-III.1} (\vec{r})$ and $z_\text{re-gal} (\vec{r})$ for \popiiione{} and galaxy-driven stages, respectively. During these active periods of ionization, each cell in our simulation is treated as either fully ionized or fully neutral.

Gas recombination is modeled during the intermediate period, $z_\text{off} < z < z_\text{flash}$, after the \popiiione{} sources have turned off and before the galaxy-driven phase begins. During this stage,  we freeze the ionization morphology at $z_\text{flash}$ by employing the \popiiione{} reionization redshift field $z_\text{re-III.1} (\vec{r})$. For all cells marked as ionized, $z < z_\text{re-III.1} (\vec{r})$, we separately reduce their ionization fraction as
\begin{align}
    x_\text{cell} (\vec{r},z)
    & = 
    \frac{\bar{x}_e(z)}{\bar{x}_\text{max}} 
    \ ,
\end{align}
when calculating \pksz{} power. This proxy treatment effectively encodes the recombination physics into the prescribed globally averaged ionization history, where all \popiiione{} sources turn off simultaneously and gas recombination is uniform in previously ionized regions.

The behavior of the ionization field due to our procedure is illustrated in Fig.~\ref{fig:time:slices}, which shows 1D field slices along a lightcone in the simulatuion grid. During the \popiiione{} stage, $z_\text{on} \geq z \geq z_\text{flash}$, ionized regions grow and merge till the field reaches maximum ionization level at $z_\text{flash}$. After this period, sources turn off and gas recombines uniformly in previously ionized regions for $z_\text{flash} > z > z_\text{off}$. At low redshifts, $z_\text{off} >  z$, reionization is driven by galaxies.

\section{Ensemble Convergence}
\label{sec:ensemble:convergence}

\begin{figure}[t!]
\centering
\includegraphics[width=\textwidth]{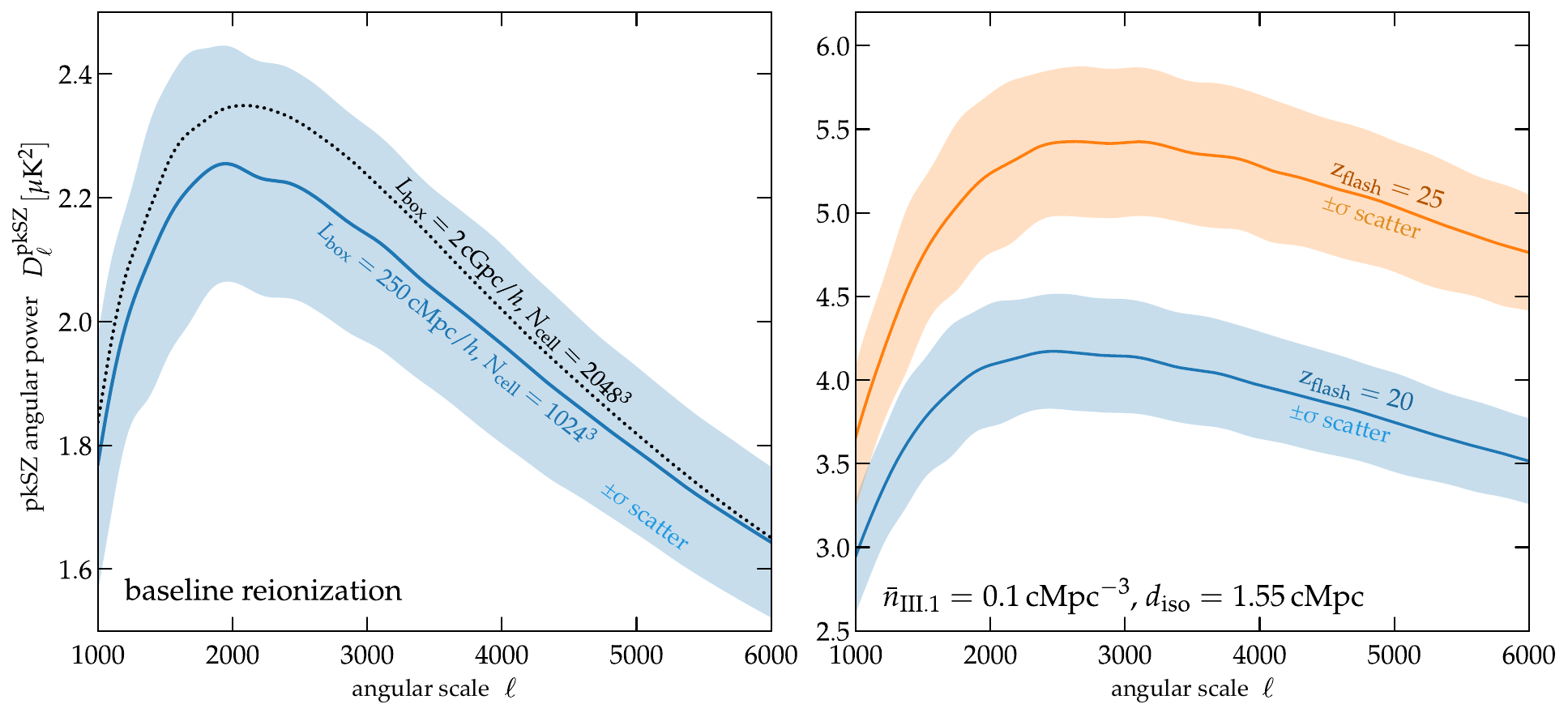}
\caption{
\pksz{} power spectra from our \amber{} ensemble runs using $L_\text{box} = 250 , \text{cMpc}/h$ and $N_\text{cell}=1024^3$. Solid lines show the ensemble mean and shaded regions indicate the corresponding $\pm 1\sigma$ scatter.
\emph{Left:} Results for the baseline reionization model, compared to a single larger-volume \amber{} realization with $L_\text{box} = 2 , \text{cGpc}/h$ and $N_\text{cell}=2048^3$ (dotted black). At $\ell=3000$, the difference between the large simulation volume and the ensemble mean is around $\sim 4.5 \%$.
\emph{Right:} Results for the $z_\text{flash}=20$ and $25$ flash ionization models using the fiducial \popiiione{} parameters, $\bar{n}_\text{III.1} = 0.1 \, \text{cMpc}^{-3}$ and $d_\text{iso} = 1.55 \, \text{cMpc}$.
}
\label{fig:amber:ensemble:behavior}
\end{figure}

The maximum \amber{} box size used in our analysis is limited by the need to resolve spatial structure in the ionization field set by the $d_\text{iso}$ prescription, as well as by computational cost. To account for finite-volume fluctuations, we generate an ensemble of 27 independent \amber{} realizations for each parameter configuration, with each realization using $L_\text{box} = 250 \, \text{cMpc}/h$ and $N_\text{cell}=1024^3$. We use the ensemble mean as the fiducial prediction in our analysis. Figure~\ref{fig:amber:ensemble:behavior} shows the ensemble mean together with the corresponding $\pm 1\sigma$ scatter in the \pksz{} power spectrum for our fiducial models. As an additional finite-volume check, we compare the ensemble mean to a single larger-volume \amber{} realization with $L_\text{box}=2 \, \text{cGpc}/h$ and $N_\text{cell}=2048^3$ for the baseline reionization model.



\bibliography{iiip1}{}
\bibliographystyle{aasjournalv7}



\end{document}